\documentclass[aps,prd,floatfix,amsmath,amssymb,preprint,eqsecnum,nofootinbib]{revtex4}
\usepackage{graphicx}
\usepackage{dcolumn}
\usepackage{bm}
\usepackage{setspace}   
\usepackage{epstopdf}
\usepackage{bbm}
\newcommand{\beq}{\begin{equation}}
\newcommand{\eeq}{\end{equation}}
\newcommand{\ba}{\begin{array}{ccc}}
\newcommand{\ea}{\end{array}}
\newcommand{\nn}{\nonumber}
 \renewcommand{\d}{\partial}
\def\bea{\begin{eqnarray}}
\def\eea{\end{eqnarray}}

\def\<{\langle}
\def\>{\rangle}
\usepackage{amsmath}
\usepackage{amssymb}
\begin{document}
\title{Wilson Loops in Non-Compact U(1) Gauge Theories at Criticality}

\author{Max~A.~Metlitski}
\email{mmetlits@fas.harvard.edu} \affiliation{Department of Physics,
Harvard University, Cambridge MA 02138, USA}

\date{\today\\[24pt]}

\begin{abstract}
We study the properties of Wilson loops in three dimensional non-compact $U(1)$ gauge theories with global 
abelian symmetries. We use duality in the continuum and on the lattice, to argue that close to the critical point between the Higgs
and Coulomb phases, all correlators of the Wilson loops are periodic functions of the Wilson loop charge, $Q$.
The period depends on the global symmetry of the theory, which
determines the magnetic flux carried by the dual particles. For single flavour scalar electrodynamics, the emergent period is $Q = 1$.
In the general case of $N$ complex scalars with a $U(1)^{N-1}$ global symmetry, the period is $Q = N$. We also give some arguments why this phenomenon does not generalize to theories with a full non-abelian $SU(N)$ symmetry, where no periodicity in $Q$ is expected. Implications for lattice simulations, as well as for physical systems, such as easy plane antiferromagnets and disordered superfluids, are noted.

\end{abstract}

\maketitle

\section{Introduction}
Three dimensional abelian gauge theories have been a subject of intense study ever since Polyakov's demonstration of monopole induced confinement in compact electrodynamics\cite{Polyakov}. The principal tool used to study confinement properties is the Wilson loop, which corresponds to an insertion of an external charge-anticharge pair into the theory. Area law for the Wilson loop indicates linear potential between distant charges and confinement.

Non-compact three dimensional $U(1)$ gauge theories with matter fields exhibit a different interesting physical phenomenon: existence of conformally invariant critical points. The prime example of such a theory is the single flavour scalar electrodynamics, the so-called Abelian Higgs model. In a certain region of parameter space, this theory exhibits a second order phase transition, which can be understood as being due to spontaneous breaking of the topological $U(1)_{\Phi}$ global flux symmetry\cite{Peskin, Halpern, Kleinert,Kovner}. The order parameter for this symmetry is the monopole operator $V(x)$, which creates a Nielsen-Oleson vortex with flux $2 \pi$. In the Higgs phase of the theory vortex excitations have a finite mass and the $U(1)_\Phi$ flux symmetry is unbroken. On the other hand, in the Coulomb phase of the theory the flux symmetry is spontaneously broken, with the photon being the corresponding goldstone. The phase transition can thus be visualized as being due to proliferation of vortices. Since the $U(1)_\Phi$ symmetry is the only global symmetry broken as one crosses the critical point, one might suspect
that the phase transition in the Abelian Higgs model is in the (inverted) $XY$ universality class and can be described in terms of a dual local theory of a dynamical vortex field $V(x)$. This hypothesis is supported by an exact duality between certain lattice versions of the Abelian Higgs and $XY$ theories\cite{Peskin,Halpern}.

The duality has been used extensively to study observables in the Abelian Higgs model, such as correlation functions of monopole, as well as magnetic field, operators. However, the behaviour of Wilson loops near the phase transition has largely escaped theoretical attention. This is not surprising: in the non-compact $U(1)$ theory there is no linear confinement between external charges, so the prime motivation for studying the behavior of Wilson loops is gone. Nevertheless, as we shall show below, absence of confinement does not preclude interesting behaviour of the Wilson loop across the phase transition.

In this paper, we discuss how to incorporate Wilson loops into the dual theory of the Abelian Higgs model using both symmetry arguments in the continuum and explicit duality transformation on the lattice. We find that a Wilson loop of charge $Q$ in the direct picture gets mapped into an infinitely thin external flux tube carrying flux $2 \pi Q$ in the dual picture\footnote{This fact has been previously noted in Ref. \cite{Samuel}, using an argument slightly different from the one presented here; however, the consequences for critical properties of the Abelian Higgs model were not discussed.}. Since flux $2 \pi$ is invisible, one immediately concludes that the universal physics near the phase transition is periodic in charge $Q$ of the Wilson line, with period $Q = 1$. This means that the behaviour of integer Wilson loops across the phase transition is non-universal: the length over which integer external charges are screened does not diverge as one approaches the critical point.
We note that the periodicity in the charge $Q$ is emergent rather than fundamental: short distance physics of scalar QED is certainly not periodic in $Q$. 

In the second part of this paper, we generalize our discussion to what we shall call the ``planar" theory: scalar QED with $N$ flavours and a $U(1)^{N-1}$ global symmetry under independent phase rotations of scalar fields, as well as a symmetry under permutations of the flavours. This theory has $N$ types of global vortices, which carry a {\it fractional} magnetic flux $2 \pi/N$\cite{Babaevfrac}. Near its critical point, the theory is believed to be dual to a theory with $N-1$ $U(1)$ gauge fields and $N$ flavours of scalars, which represent global vortices of the direct model\cite{mv,SachdevReview,Sudbo}. We show that due to the fractional charge of the vortices under the flux symmetry, a Wilson line of charge $Q$ in the direct theory gets mapped to an external magnetic flux tube with flux $2\pi Q/N$ in the dual theory. Thus, near the critical point, the physics of the $N$ flavour model is periodic in the charge $Q$ of the Wilson line with period $Q = N$.

The emergent periodicity in the charge $Q$ can be explicitly tested by lattice simulations. We suggest one observable for lattice simulations: the electric field produced by a straight temporal Wilson line at the phase transition. Like all universal observables, the electric field will be periodic in the charge $Q$ of the Wilson line. We also present some explicit semi-quantitative predictions for the coefficient of the electric field based on the $1/M$ expansion of the dual theory.

Finally, we would like to understand whether the phenomena described above generalize to $U(1)$ gauge theory with $N$ scalar flavours and a full $SU(N)$ global symmetry. We shall argue that the answer to this question is no: the physics in the $SU(N)$ symmetric model is not periodic in the charge $Q$ of the Wilson line. In contrast to the situation in the theory with abelian global symmetry, the candidate dual degrees of freedom in the $SU(N)$ symmetric model cannot be associated with {\it local} fields charged under the flux symmetry. As a consequence, no periodicity in the charge $Q$ emerges.

We would like to note that besides being of general theoretical interest, the behaviour of Wilson loops across phase transitions in non-compact $U(1)$ gauge theories is important for a number of physical problems. For instance, it is believed that the phase transition from an antiferromagnetic N\'eel state to a valence bond solid (VBS) state on a square lattice is described by $N = 2$ scalar non-compact QED\cite{senthil1,senthil2}. The model with $SU(2)$ flavour symmetry corresponds to a spin-rotation invariant system, while the model with an abelian $U(1)$ global symmetry describes, the so-called, easy-plane antiferromagnet. The later model is also believed to describe the phase transition from a superfluid state to a disordered state in a theory of lattice bosons\cite{balents1}. A missing spin impurity in this class of models is represented by a Wilson line in the gauge-theory description\cite{kolezhuk}. We have applied the dual description of the Wilson loops presented in this paper to study impurity induced VBS susceptibility in easy-plane antiferromagnets in Ref.\cite{MSVBS}. The predictions of \cite{MSVBS} may be explicitly tested by lattice studies of phase transitions in antiferromagnets and possibly by STM experiments on cuprate compounds.

This paper is organized as follows. In section \ref{sec:AHM} we describe how to incorporate Wilson loops into a dual description of the Abelian Higgs model. In section, \ref{sec:EasyPlane} we generalize this description to a model with $N$ flavours and a $U(1)^{N-1}$ global symmetry. A discussion of the curious periodicity in the charge $Q$ of the Wilson loop appears in section \ref{sec:Discussion}. Section \ref{sec:SUN} contrasts the behaviour of Wilson loops in theories with abelian and non-abelian global symmetry.  Concluding remarks are presented in section \ref{sec:Conclusion}.

\section{Wilson Loops in the Abelian Higgs Model}\label{sec:AHM}
\subsection{Duality and Wilson loops}\label{sec:DualityW}
It is well known that in three space-time dimensions, near its critical point, non-compact $N=1$ scalar electrodynamics is dual to a theory of a complex (pseudo)scalar field with a global $U(1)$ symmetry \cite{Peskin,Halpern,Kleinert,Kovner}. The Lagrangians of these two theories are as follows,
\bea \label{LQED}L_{QED} &=& \frac{1}{2 e^2} F^2_{\mu} + |(\d_{\mu} - i A_{\mu}) z|^2 + m^2 |z|^2 + \frac{g}{2} |z|^4 \\
\label{LXY}L_{XY} &=& |\d_{\mu} V|^2 + \tilde{m}^2 |V|^2 + \frac{\tilde g}{2} |V|^4\eea
Here $A_{\mu}$ is a non-compact gauge field, $F_{\mu} = \epsilon_{\mu \nu \lambda} \d_{\mu} A_{\lambda}$ is the magnetic field, and  $z$ and $V$ are complex one component fields. The duality is understood as being true for the range of parameters where $L_{QED}$ has a second order phase transition (which at weak coupling is believed to occur for $g/e^2$ sufficiently large). One way to understand the duality is by noting that the phase transition in scalar QED is driven by spontaneous breaking of flux symmetry $U(1)_\Phi$, which is precisely the global symmetry of $L_{XY}$. The order parameter for breaking of the flux symmetry is the monopole operator $V(x)$ - that is the dynamical field of $L_{XY}$. As we know, to each continuous symmetry there corresponds a conserved current. In the case of flux symmetry of QED, this pseudo-vector current is just the magnetic field $F_{\mu}$, which is trivially conserved in the absence of monopoles, $\d_{\mu} F_{\mu} = 0$. Let's introduce an external field $H_{\mu}$ that would couple to this current,
\beq \delta L_{QED} = i H_{\mu} F_{\mu}\eeq  Suppose we are calculating some correlation function with insertion of a string of monopole operators $\{V^{q_i}(x_i)\}$ of charge $q_i$ at points $x_i$. The gauge field $A_{\mu}$ in the path integral is then subject to the condition, $\d_{\mu} F_{\mu} = \sum_i 2\pi q_i \delta(x-x_i)$. Then under the transformation,
\beq \label{Htransf} H_\mu \to H_{\mu} + \d_{\mu} \alpha\eeq
\beq S_{QED} \to S_{QED} + i \int dx \,\d_{\mu} \alpha F_{\mu} = S_{QED} - i \int dx\, \alpha \,\d_{\mu} F_{\mu} = S_{QED} - 2 \pi i \sum_i q_i \alpha(x_i)\eeq
Hence, by introducing the field $H_{\mu}$ we can enlarge the global $U(1)_\Phi$ symmetry to a fictitious local symmetry, provided that the monopole operators transform as,
\beq \label{Vqtransf} V^{q}(x) \to e^{2 \pi i q \alpha(x)} V^q(x)\eeq 
The dual Lagrangian $L_{XY}$ has to posses this local symmetry. Hence, to introduce the field $H_{\mu}$ into the dual Lagrangian we simply have to covariantize the derivative of the dynamical monopole field $V$,
\beq \label{covar} \d_{\mu} V \to D_{\mu} V = (\d_{\mu} - 2 \pi i H_{\mu}) V\eeq in eq. (\ref{LXY}). Other ``gauge invariant" operators can also be added to $L_{XY}$, e.g. $H^2_{\mu \nu}$; however, their contribution will, generally, either cancel out in correlation functions or be less singular near the critical point (see section \ref{sec:pertloc} for a more detailed discussion).

Thus, the dual Lagrangian in the presence of a background source field $H_{\mu}$ is given by,
\beq \label{dualH} L_{XY} = |(\d_{\mu}-2 \pi i H_{\mu}) V|^2 + \tilde{m}^2 |V|^2 + \frac{\tilde g}{2} |V|^4\eeq

The covariantization procedure (\ref{covar}) was explicitly written down in Ref.~\onlinecite{Laine}. Similar arguments for the case of a constant imaginary $H_{\mu}$, which physically represents an external magnetic field in the QED language and translates into a chemical potential for the flux symmetry in the XY language, have been given in Ref.~\onlinecite{Son}. In the next section, we shall also give an argument based on an exact duality transformation on the lattice, which will support (\ref{dualH}).

Having learned how to incorporate the source field $H_{\mu}$ into the dual Lagrangian, it is now trivial to dualize Wilson loops. Indeed, insertion of a Wilson loop $W({\cal C})$ into a correlation function is equivalent to adding into the Lagrangian the source term
\beq \delta L = i Q \int_{\cal C} dx_{\mu} A_{\mu} = i Q \int_{\cal S} dS_{\mu} F_{\mu} = i \int dx H_{\mu} F_{\mu} \eeq
where the surface ${\cal S}$ satisfies, $\d {\cal S} = {\cal C}$ and
\beq \label{HWilson} H_{\mu}(x) = Q \int_{y \in {\cal S}} dS_\mu \,\delta(x-y)\eeq So $H_{\mu}$ is a field that lives on the surface of the Wilson loop and is directed perpendicular to this surface. 

Another benefit of introducing the source field $H_{\mu}$ is that by differentiating with respect to it we can compute correlation functions of the magnetic field $F_{\mu}$. For instance,
\beq \langle - i F_{\mu} (x) \rangle_H = \frac{\delta \log Z[H]}{\delta H_{\mu}(x)}=-2 \pi i \langle \left(V^{\dagger} D_{\mu} V - (D_{\mu} V)^{\dagger} V\right)(x)\rangle_H\eeq
Hence the topological flux current $F_{\mu}$ of QED gets mapped into the Noether's current associated with the global $U(1)$ symmetry of the dual model. 

We have seen above that the Wilson loop $W({\cal C})$ in the dual theory is specified by a surface ${\cal S}$ with $\d {\cal S} = {\cal C}$ rather than by the contour ${\cal C}$ alone. Let's investigate the dependence of the dual theory on the choice of this surface.
If we pick a different surface ${\cal S}'$, with $\d{\cal S'} ={\cal C}$ then the field $H_{\mu}$ undergoes a gauge transformation $H_{\mu} \to H'_{\mu} = H_{\mu} + \d_{\mu} \alpha$ with $\alpha(x) = -Q\, 1_{x \in {\cal V}}$ where ${\cal V}$ is the volume bounded by the two surfaces ${\cal S}$ and ${\cal S}'$. Hence,
\beq \langle V(x) ... \rangle_{H'} = e^{2 \pi i \alpha(x)} \langle V(x) ... \rangle_{H}\eeq
where ellipses denote some other operators. Thus, the operator $V(x)$ is invariant under changing the surface of the Wilson loop if and only if $Q$ is an integer. This is nothing but Dirac's condition expressed in the language of the dual theory.
However, a theory with arbitrary non-integer $Q$ is still sensible provided that we don't consider monopole operator insertions, or more formally, confine our attention to correlation functions of operators which are invariant under the fictitious $U(1)_{\Phi}$ local symmetry, e.g. the magnetic field operator $-i F_{\mu} = - 2 \pi i V^{\dagger}\overleftrightarrow{D}_{\mu} V$. In fact, if we are dealing with such gauge invariant operators we don't necessarily have to use the precise form of $H$ given by (\ref{HWilson}); defining $\gamma_{\mu}$ to be a field living on the perimeter of the Wilson loop and directed along it, 
\beq \gamma_{\mu}(x) = Q \int_{y\in {\cal C}} dy_{\mu} \delta(x-y)\eeq we see that,
\beq \label{curlH} \epsilon_{\mu \nu \lambda} \d_{\nu} H_{\lambda} = \gamma_{\mu}\eeq
Then, by performing a suitable gauge transformation on $H_{\mu}$ and $V$ we can choose $H_{\mu}$ to be any field with curl given by $\gamma_{\mu}$. Thus, we see that the duality maps a Wilson loop of charge $Q$ in the QED language to an external magnetic flux tube of flux $2 \pi Q$ in the XY language. This correspondence has been noted in Ref.~\onlinecite{Samuel}, but the consequences of this correspondence for the critical properties of Wilson loops were not discussed.


Thus, we have to solve an Aharonov-Bohm like problem for the dual field $V$. The question is simplest to analyze with the original gauge choice (\ref{HWilson}). This ``string" gauge is equivalent to $H_{\mu} = 0$ and the boundary condition,
\beq \label{bc} V(x^+) = e^{2 \pi i Q} V(x^-)\quad \mathrm{for }\,\, x \in {\cal S} \eeq
where $x^\pm$ denote points on opposite sides of the surface ${\cal S}$ ($x^\pm = x \pm \epsilon n$, for $\epsilon \to 0^+$, where $n$ is a local normal to ${\cal S}$). So the Wilson loop imposes a twisted boundary condition (\ref{bc}) in the dual theory.
%
We observe that the physics is, therefore, a periodic function of $Q$. For integer $Q$ the boundary condition (\ref{bc}) is trivial - there is no twist. So our argument indicates that integral Wilson lines do not affect the physics on distances of order of the correlation length of the theory: screening of integral charges takes place on length scale which does not diverge as one approaches the phase transition. 
This is certainly an unexpected result: we will discuss it further in section \ref{sec:Discussion}. However, first we would like to obtain further support for this result by performing an explicit duality on the lattice, which will provide additional physical insight into the origin of the periodicity in $Q$.

Another interesting consequence of periodicity in $Q$ is the emergence of charge conjugation $C$ and charge parity $CP$ invariance at points $Q = \pm 1/2$. Indeed, the Wilson line generally breaks both $C$ and $CP$ symmetries, which map the charge of the Wilson line $Q \to -Q$. However,  due to periodicity in $Q$, the points $Q = \pm 1/2$ are identified, so $C$ and $CP$ symmetries are effectively restored for half-integer valued $Q$.


\subsection{Duality on the Lattice}\label{sec:LatticeDual}

In Section \ref{sec:DualityW} we have given arguments on how to perform the duality on
$\mathrm{QED}_3$ with Wilson loop insertions. Our arguments were very general, being based on the
presence of flux-symmetry alone. In the present section, we would like to support the arguments of \ref{sec:DualityW}
by performing an exact duality between lattice versions of QED and XY model. In the process, we will obtain some insight
into the emergent periodicity in the charge of the Wilson line.

The lattice duality between non-compact QED and XY model is very well known \cite{Peskin,Halpern}. We start from the QED lattice action,
\beq \label{SQEDLat} S_{QED} = \frac{1}{2 e^2} \sum_{\bar{j} \mu} (\Box A)_{\bar{j} \mu}^2 + \frac{1}{2 g} \sum_{j \mu} (d \theta - A - 2 \pi n)^2_{j \mu}\eeq
Here $A_{j \mu}$ is a gauge field living on links of the lattice and $e^{i \theta}$ is a matter field, whose amplitude is frozen. The auxillary variables $n_{j \mu}$ are integers living on the links of the lattice, whose purpose is to ensure the  $2 \pi$ periodicity of the variable $\theta$. For our purposes it will also be useful to add sources corresponding to monopole  and Wilson loop insertions into the action,
\beq \label{SQEDSource} S_{QED} = \frac{1}{2 e^2} \sum_{\bar{j} \mu} (\Box A - 2 \pi r)_{\bar{j} \mu}^2 + \frac{1}{2 g} \sum_{j \mu} (d \theta - A - 2 \pi n)^2_{j \mu} + i \sum_{\bar{j}\mu} H_{\bar{j} \mu}(\Box A - 2 \pi r)_{\bar{j}\mu}\eeq
Here $r_{\bar{j}\mu}$ is an integer valued source field representing Dirac strings running from locations of monopoles to infinity, satisfying,
\beq \label{divr}(\nabla \cdot r)_{\bar{j}} = -s_{\bar j} = -\sum_i q_i \delta_{{\bar j} {\bar j}_i}\eeq
where $q_i$ and $\bar{j}_i$ are correspondingly charges and locations of monopole insertions. The field $H_{\bar{j} \mu}$ in (\ref{SQEDSource}) is a source coupling to the physical magnetic field $(\Box A - 2 \pi r)_{\bar{j}\mu}$ (i.e. the magnetic field with the Dirac string subtracted). To represent a Wilson loop of charge $Q$, we can choose $H_{\bar{j} \mu}$ to be equal to $Q$ on the surface perpendicular to the loop and zero everywhere else. For integer $Q$, such a Wilson loop will be independent of the choice of the surface, while for non-integer $Q$ it will depend on the choice of the surface if monopole operator insertions are present.

Now, we perform the duality. First, we decouple the kinetic term for the gauge field by introducing an auxiliary field $P_{\bar{j} \mu}$.  We also Poisson resum the field $n_{j \mu}$ by introducing an integer valued variable $J_{j \mu}$.
\beq S_{QED} \to \frac{e^2}{2} \sum_{\bar{j} \mu} P^2_{\bar{j} \mu} + \frac{1}{2 g} \sum_{j \mu} (d \theta - A - 2 \pi n)^2_{j \mu} + 2 \pi i \sum_{j \mu} n_{j \mu} J_{j \mu} + i \sum_{\bar{j} \mu} (P+H)_{\bar{j} \mu} (\Box A - 2 \pi r)_{\bar{j} \mu}\eeq
After Poisson resummation, $n_{j \mu}$ becomes a free real variable, so we can shift it, $2 \pi n'_{\bar{j}\mu} = ( 2\pi n + A -d \theta)_{\bar{j} \mu}$,
\beq S_{QED} \to \frac{e^2}{2} \sum_{\bar{j} \mu} P^2_{\bar{j} \mu} + \frac{1}{2 g}\sum_{j \mu} (2 \pi n')^2_{j \mu} + i \sum_{j \mu} (2 \pi n'-A+d \theta)_{j \mu} J_{j \mu} + i \sum_{\bar{j} \mu} (P + H)_{\bar{j} \mu}(\Box A - 2 \pi r)_{\bar{j} \mu}\eeq
Now, performing the integral over $\theta_{j \mu}$ we obtain a constraint 
\beq \label{mconstr} \nabla \cdot J = 0\eeq Physically, $J_{j \mu}$ represents the world-lines of $e^{i \theta}$ particles. If we were to integrate over all other fields in the problem, we would see that these worldlines interact with a long-range $1/r$ Coulomb interaction (there are also local interactions between these worldlines controlled by the coupling strength $g$). We solve the constraint (\ref{mconstr}) in terms of an integer valued field $b_{\bar j \mu}$,
\beq \label{mconstsol} J_{j \mu} = (\Box b)_{j \mu}\eeq Now, we rearrange our action slightly and integrate over the $n'$ field,
\beq S_{QED} \to \frac{e^2}{2} \sum_{\bar{j} \mu} P^2_{\bar{j} \mu} + i \sum_{\bar{j} \mu} \left(\Box(P+H) - J\right)_{j \mu} A_{j \mu} - 2 \pi i \sum_{\bar{j} \mu} (P+H)_{\bar{j} \mu} r_{\bar{j} \mu} + \frac{g}{2} \sum_{j \mu} J^2_{j \mu}\eeq
Performing the integral over the $A$ field, we obtain a constraint,
\beq \label{Pconstr}\Box(P+H)_{j \mu} - J_{j \mu} = 0\eeq
Recalling (\ref{mconstsol}) we can solve (\ref{Pconstr}) by introducing a real field $\varphi$,
\beq P + H - b = \frac{d \varphi}{2 \pi}\eeq
arriving at the action,
\beq S_{QED} \to \frac{e^2}{8 \pi^2} \sum_{{\bar j} \mu}(d \varphi - 2 \pi H + 2 \pi b)^2_{\bar{j} \mu} + \frac{g}{2} \sum_{j \mu} (\Box b)^2_{j \mu}-i \sum_{\bar{j} \mu} (d \varphi + 2\pi b)_{\bar{j} \mu} r_{\bar{j} \mu}\eeq
Recalling that both $b$ and $r$ are integer valued, summing the last term by parts and using (\ref{divr}),
\beq \label{Sdual1} S_{QED} \to \frac{e^2}{8 \pi^2} \sum_{{\bar j} \mu}(d \varphi - 2 \pi H + 2 \pi b)^2_{\bar{j} \mu} + \frac{g}{2} \sum_{j \mu} (\Box b)^2_{j \mu} - i \sum_{\bar{j}} \varphi_{\bar{j}} s_{\bar{j}}\eeq
Eq. (\ref{Sdual1}) is the final form of the dual lattice action. Temporarily setting $g = 0$, dropping all the source field, we obtain the usual Villain form of the XY model, with $e^{i \varphi}$ being the XY field. As already noted, the role of finite $g$ is to introduce short-range interactions between $e^{i \theta}$ particles, that is between vortices of the $\varphi$ field. This is also evident from the dual action (\ref{Sdual1}) as we can identify $\Box b$ with the density of $\varphi$ vortices. Since such vortices for $e^2 \neq 0$ already interact with a long-range Coulomb potential, we expect the phase transition at finite $g$ to be in the same (inverted) XY universality class as at $g = 0$.

Now, restoring the source fields into (\ref{Sdual1}), we immediately identify $e^{i \varphi}$, with the monopole field of QED. Moreover, we also see that the source field $H$ enters the dual action by gauging the lattice derivative of the $\varphi$ field. We have predicted this fact from symmetry arguments alone in Section \ref{sec:DualityW}. Hence, we see that the action (\ref{dualH}) is a suitable continuum generalization of our dual action (\ref{Sdual1}), where we identify $V \sim e^{i \varphi}$. 

Now, we come to the question that interests us most: what is the influence of integral Wilson loops on our theory. For a general non-integral charge $Q$ the Wilson loop enters the dual theory as a highly non-local coupling, as the source field $H$ lives on the whole surface of the Wilson loop, rather than on its perimeter. Nevertheless, for an integral charge $Q$, we can perform a transformation,
\beq b' = b - H\eeq
as in this case $H$ is integer valued. Hence, dropping the monopole insertions,
\bea S_{XY} &=& \frac{e^2}{8 \pi^2} \sum_{{\bar j} \mu}(d \varphi  + 2 \pi b')^2_{\bar{j} \mu} + \frac{g}{2} \sum_{j \mu} (\Box (b'+H))^2_{j \mu}\\&=&\frac{e^2}{8 \pi^2} \sum_{{\bar j} \mu}(d \varphi  + 2 \pi b')^2_{\bar{j} \mu} + \frac{g}{2} \sum_{j \mu} (\Box b'+\gamma)^2_{j \mu}\label{SXYgamma}\eea
where we used $(\Box H)_{j \mu} = \gamma_{j \mu}$ with the $Q$ valued vector field $\gamma$ pointing along the perimeter of the Wilson loop (and being zero everywhere else). Thus, for integral $Q$ the coupling of the theory to the Wilson loop becomes local. So we expect the physics of integer Wilson loops to be drastically different from that of non-integer ones.

Moreover, we see that for the special value $g = 0$, the coupling to the Wilson loop in (\ref{SXYgamma}) disappears all together. From the point of view of the direct theory (\ref{SQEDLat}) this fact is not surprising: setting $g = 0$ forces $A = d\theta - 2\pi n$, so that $W({\cal C}) = 1$.  Thus, the theory possesses a limit in which the insertion of integer Wilson loops is trivial, but as one varies $e^2$, a phase transition in the (inverted) XY universality class still occurs. As already noted, we expect the phase transition at $g \neq 0$ to be in the same universality class as that of $g = 0$, that is $g$ is in some sense an ``irrelevant" coupling.  Hence, one can argue that the physics of integer Wilson loops is non-universal - their only effects appear on distance scale of order of the microscopic cutoff of the theory. 

To illuminate this conclusion further, let us rewrite the dual theory (\ref{Sdual1}) in terms of magnetic vortex lines, (i.e. worldlines of the field $e^{i \varphi}$). We proceed by Poisson resumming the variable $b_{\bar{j} \mu}$ by introducing an integer valued field $l_{\bar{j} \mu}$,
\beq S_{QED} = \frac{e^2}{8 \pi^2} \sum_{\bar{j} \mu} (d \varphi - 2 \pi H + 2 \pi b)_{\bar{j} \mu}^2 + \frac{g}{2}\sum_{j \mu} (\Box b)^2_{j \mu} + 2 \pi i \sum_{\bar{j} \mu} l_{\bar{j} \mu} b_{\bar{j} \mu}\eeq
where we've dropped the monopole source $s$ for simplicity. Now, shifting $2 \pi b' = 2 \pi b - 2 \pi H + d\varphi$,
\beq S_{QED} \to \frac{e^2}{2} \sum_{\bar{j} \mu} b'^2_{\bar{j} \mu} + \frac{g}{2}\sum_{j \mu} (\Box b' + \gamma)^2_{j \mu} + i \sum_{\bar{j} \mu} l_{\bar{j} \mu} (2 \pi  b' + 2 \pi H - d \varphi)_{\bar{j} \mu}\eeq
Integrating over the field $\varphi$, we obtain the constraint,
\beq \nabla \cdot l = 0\eeq
Physically, $l_{j \mu}$ are just the magnetic flux tubes. We can also integrate over the field $b'$, obtaining,
\beq \label{SdualL}S_{QED} = \frac{1}{2} \sum_{j j' \mu} (2 \pi l)_{\bar{j} \mu} D_{jj'}(2 \pi l)_{\bar{j'} \mu} + 2 \pi i \sum_{j j' \mu} H_{\bar{j} \mu} (e^2 D_{j j'}) l_{\bar{j'} \mu}  + \frac{e^2 g}{2} \sum_{j j'\mu} \gamma_{j \mu} D_{j j'} \gamma_{j' \mu}\eeq
where the propagator $D_{j j'}$ is given by,
\beq D_{j j'} = \frac{1}{V} \sum_k \frac{1}{e^2+g\sum_{\nu} 4 \sin^2{\frac{k_{\nu} a}{2}}}e^{i k (j-j')}\eeq
with $a$ being the lattice spacing and $V$-the number of sites in the lattice. The last term in (\ref{SdualL}) does not couple to the dynamical field $l$ and, thus, is trivial. The first term in (\ref{SdualL}) represents the short range interaction between the flux-tubes. The second term,
\beq \label{Fluxl} i \Phi =  2 \pi i \sum_{j j' \mu} H_{\bar{j} \mu} (e^2 D_{j j'}) l_{\bar{j'} \mu}\eeq
is the one that interests us in conjunction with the properties of Wilson loops. Indeed, we identify,
\beq B_{\bar{j} \mu} = 2 \pi \sum_{j'} (e^2 D_{j j'}) l_{\bar{j'} \mu}\eeq
with the magnetic field produced by each flux tube. We see that the magnetic field due to each flux-line is short-range, as expected. Hence, if we take $H$ to represent the Wilson loop, (\ref{Fluxl}) simply adds up the contribution of all flux tubes to the flux through the loop. In the special limit $g = 0$, the propagator $D_{j j'}$ is ultra-local, $e^2 D_{j j'} = \delta_{j j'}$ and the flux-lines become infinitely thin. Then, (\ref{Fluxl}) simply counts the number $n$ of flux-lines passing through the loop (in other words, $n$ is the linking number of the flux-tubes with the Wilson loop),
\beq \label{linking1} i \Phi \to 2 \pi i Q n\eeq
We immediately see that for $Q$ - integer, the term (\ref{linking1}) gives a trivial contribution. 

Now, turning $g$ back on, our flux-lines obtain a finite thickness $r_f$. Then the expression for the flux (\ref{linking1}) is modified by contributions from flux-tubes passing within a distance $\sim r_f$ from the boundary of the Wilson loop (that is flux-lines, which are not entirely inside or outside the loop). This contribution can be understood as a local coupling to the Wilson line, which is expected to be less relevant in the RG sense than the non-local term (\ref{linking1}). In fact, in the next section, we will argue that all such local, line-like coupling terms are irrelevant.


\subsection{Perturbations local at the Wilson line}\label{sec:pertloc}
Another way to argue the non-universality of response to integral external charges is to try to construct a relevant perturbation of the continuum dual theory (\ref{dualH}). Such perturbations have to be invariant under the fictitious $U(1)_\Phi$ local symmetry discussed in Section \ref{sec:DualityW}. Moreover, as discussed in section \ref{sec:LatticeDual}, for integral $Q$, the perturbation must be local to the Wilson line. There are plenty of operators with correct symmetry properties, since in the action (\ref{dualH}) we included only the most relevant (in terms of power counting) terms. We count the source field $H_{\mu}$ as having dimension $1$. This is the canonical dimension of this field. Taking into account $C$ and $P$ symmetry, the operators of lowest dimension which we can write down are,
\bea &&\left[(\epsilon_{\mu \nu \lambda} \d_{\nu} H_{\lambda})^2\right]  = \left[\gamma^2_{\mu}\right]= 4\label{Hstrength}\\
&&\label{gammasqr}\left[(\epsilon_{\mu \nu \lambda} \d_{\nu} H_{\lambda})^2 V^{\dagger} V\right] = \left[\gamma^2_{\mu} V^{\dagger} V\right] = D+2\\
&&\left[(-i \epsilon_{\mu \nu \lambda} D_{\nu} V^{\dagger} D_{\lambda} V) (\epsilon_{\mu \alpha \beta} \d_{\alpha} H_{\beta}) \right] = \left[(-i \epsilon_{\mu \nu \lambda} D_{\nu} V^{\dagger} D_{\lambda} V) \gamma_{\mu}\right] = D+2 \label{vortexdens}
\eea
where we've included the canonical dimensions of the operators in question. The first operator (\ref{Hstrength}) does not couple to the dynamical fields of the dual theory and, therefore, its contribution is trivial. The other operators are irrelevant by power-counting. We  particularly want to draw attention to the operator (\ref{vortexdens}), since it couples the Wilson line $\gamma_{\mu}$ to the vortex density operator, $-i \epsilon_{\mu \nu \lambda} D_{\nu} V^{\dagger} D_{\lambda} V$. This is precisely the kind of coupling that we have for integral charges in the dual lattice action (\ref{SXYgamma}), namely $\Box b'\, \gamma$. 

We note that the above argument might be too naive, as the field strength $\gamma_{\mu} = \epsilon_{\mu \nu \lambda} \d_{\nu} H_{\lambda} \sim \delta^2(\vec{x})$ corresponding to the Wilson line is very singular. For instance, it is not clear how to interpret  the $\gamma^2_{\mu}$ term in (\ref{gammasqr}). If one takes $\gamma^2_{\mu} \sim (\delta^2(\vec{x}))^2 \sim \Lambda^2 \delta^2(\vec{x})$, then the extra factor of $\Lambda^2$ upsets our power counting scheme. 
Such singularities will occur at all orders in field-strength. So perhaps it is more appropriate for integer-valued external charge to simply write all operators in the dual theory, which live locally on the Wilson line (of course, we then don't know the dependence of the coefficients of these operators on charge $Q$). The operator which would seem to be most relevant is,
\beq \label{VVWilson} \delta L = u \int d\tau \,V^{\dagger} V(\vec{x} = 0, \tau)\eeq  Here $u$ is some real coupling constant,
and at the critical point its scaling dimension is determined following Ref. \cite{qimp1}, 
$\mbox{dim}[u] = 1 - \mbox{dim}[V^\dagger V] = 1 - (3 - 1/\nu) = 1/\nu - 2$. Because $\nu \approx 2/3$ for the XY model, $u$ is an irrelevant perturbation.
We, therefore, come to the curious conclusion that it is impossible to construct a (weak) perturbation localized on a one-dimensional line, which would be relevant at the fixed point of the XY model.

\section{Wilson Loops in $N$-flavour Model with $U(1)^{N-1}$ symmetry}\label{sec:EasyPlane}

In this section, we consider a theory with $N$ flavours of scalar fields $z_{\alpha}$ ($N$ does not necessarily have to be large),
\beq \label{QEDN} L = \frac{1}{2 e^2} F^2_{\mu} + |(\d_{\mu}-i A_{\mu})z_{\alpha}|^2 + U(z_{\alpha})\eeq
We refer to this theory as the ``planar" model. Here, $U$ is some potential with the global $U(1)^{N}$ symmetry under independent phase rotations of the $z_{\alpha}$ fields.
The singlet component of this symmetry is actually gauged by the field $A_{\mu}$,
\beq\label{loc} U(1):\, z_{\alpha} \to e^{i \theta(x)} z_{\alpha}, \quad A_{\mu} \to A_{\mu} + \d_{\mu} \theta\eeq
while the non-singlet components are true global symmetries of the theory,
\beq\label{glob} U(1)^{N-1}:\, z_{\alpha} \to e^{i \theta^a t^a_{\alpha}} z_{\alpha}\eeq
where $t^a$, $a = 1 .. N-1$ are the generators of the $U(1)^{N-1}$ symmetry satisfying, $\sum_{\alpha} t^a_{\alpha} = 0$.
We require $U$ to
have a symmetry under the permutation of labels of $z_{\alpha}$ fields.  We choose $U$ in such a fashion that in the ``condensed" phase of the theory, it favours non-zero expectation values of all components of the $z_{\alpha}$ field, so that the vacuum manifold of the theory is a torus, $(S^1)^N$ (here we temporarily forget that the singlet symmetry is gauged). For $N = 2$ the theory under consideration is believed to describe the phase transition in the easy-plane antiferromagnet. 

We would like to dualize the theory (\ref{QEDN}). Similar theories were dualized in Refs.~\onlinecite{mv,SachdevReview,Sudbo}, and here we will
present a related discussion aimed at incorporating Wilson loops into the dual theory. An exact duality on the lattice appears in the appendix, but we can write down the form of the dual action from very general considerations. Let us first identify the dual degrees of freedom. We go to the condensed phase of the theory (\ref{QEDN}), where all $\langle z_{\alpha} \rangle \neq 0$. Then, we can have vortices in any component of the $z_{\alpha}$ field. Formally, the homotopy group, $\pi_1((S^1)^N) = {\mathbb Z}^N$. So, we have $N$ types of vortices, which become the degrees of freedom of the dual theory $V_{\alpha}$, $\alpha = 1..N$. 

These vortices are global, rather than local. Indeed, let's consider a vortex in the first component $z_1$,
\beq z_1(\vec{x}) \sim v e^{i \lambda(\vec{x})}, \quad z_{\alpha} \sim v,\, \alpha \neq 1, \quad |\vec{x}| \to \infty\eeq
where $\lambda(\vec{x})$ winds from $0$ to $2 \pi$ as one goes around a contour out at infinity surrounding the vortex. Then, this vortex corresponds to a space-time dependent transformation of the vacuum (\ref{loc}), (\ref{glob}), with, $\theta(\vec{x}) = \frac{1}{N} \lambda(\vec{x})$ and $\theta^a(\vec{x}) t^a = (1-1/N,-1/N,...-1/N) \lambda(\vec{x})$. Thus, our vortex possesses a winding both in the local and in the global symmetry group. The winding in the local $U(1)$ group will be canceled by the gauge field,
\beq A_{\mu}(x) = \d_{\mu} \theta(x) = \frac{1}{N} \d_{\mu} \lambda(x)\eeq
hence our global vortices carry a magnetic flux $\Phi = 2 \pi/N$. Therefore, under the flux symmetry (\ref{Htransf}), the fields $V_{\alpha}$ should transform as,
\beq \label{VNtransf} V_{\alpha}(x) \to e^{2 \pi i \alpha(x)/N} V_{\alpha}(x)\eeq 
This fact will be crucial for the analysis to follow. 

The winding in the global group will lead to a long-range Coulombic interaction between our vortices. We will need dynamical gauge fields in the dual theory to give rise to this interaction. However, if we have a unit winding in each component of the $z$ field, our vortex becomes completely local, and carries total flux $2 \pi$. We can think of such a local vortex as a composite of $N$ global vortices of different types. The creation operator for this flux-tube, therefore, will be,
\beq \label{VN} {\cal V}(x) = \prod_{\alpha} V_{\alpha}(x)\eeq
Since the local vortex carries flux $2 \pi$, we can also associate the operator (\ref{VN}) with the monopole operator of the direct theory. Indeed, given (\ref{VNtransf}), under the flux symmetry (\ref{Htransf}),
\beq {\cal V}(x) \to e^{2 \pi i \alpha(x)} {\cal V}(x) \eeq
which is the correct transformation law for the monopole operator (\ref{Vqtransf}). 

We expect local vortices to interact by short range forces. Therefore, the operator (\ref{VN}) should not be charged under the emergent gauge fields of the dual theory. 

We are now ready to write down the dual theory,
\beq \label{dualN} L = \frac{1}{2 \tilde{e}^2} \sum_i (F^{\alpha}_{\mu})^2 +  |(\d_{\mu} - i B^{\alpha}_{\mu} - \frac{2 \pi i}{N} H_{\mu})V_{\alpha}|^2 + \tilde{U}(V_{\alpha})\eeq
Here $B^\alpha_{\mu} = B^{a}_{\mu} t^a_{\alpha}$, $a = 1..N-1$, are emergent dual gauge fields, which couple to the non-singlet currents. $F^{\alpha} = \epsilon_{\mu \nu \lambda} \d_{\nu} B^{\alpha}_{\lambda}$ are the corresponding field strengths. The dual potential $\tilde{U}(V_{\alpha})$ is chosen to have the same properties as the direct potential $U$: it has a $U(1)^N$ symmetry under independent phase rotations of the fields $V_{\alpha}$ and a symmetry under permutation of labels of $V_{\alpha}$ fields. Moreover, it favours $\langle V_{\alpha} \rangle \neq 0$ for all $\alpha$ in the condensed phase of the dual theory. Thus, the theory (\ref{dualN}) has a local $U(1)^{N-1}$ symmetry,
\beq \label{U1Nloc} U(1)^{N-1}: \quad V_\alpha(x) \to e^{i \phi^a(x) t^a_{\alpha}} V_{\alpha}(x), \quad B^a_{\mu} \to B^a_{\mu} + \d_{\mu}\phi^a\eeq
as well as the global $U(1)$ flux symmetry of the direct theory (\ref{VNtransf}) (which we have promoted to a local symmetry by introducing a non-dynamical source field $H_{\mu}$). As required, the monopole operator (\ref{VN}) is invariant under the local $U(1)^{N-1}$ symmetry of the dual theory (\ref{U1Nloc}).

The theory (\ref{dualN}) also has a global $U(1)^{N-1}$ symmetry associated with conservation of fluxes of the $N-1$ emergent gauge fields. This topological symmetry can be identified with the Noether's symmetry (\ref{glob}) of the direct theory.

Now, we would like to apply the duality discussed above to study the properties of Wilson loops. Recall, that to represent Wilson loops we must use a source field $H_{\mu}$ given by (\ref{HWilson}). As discussed for the case of $N=1$ theory, the effect of such a source field on the dual action (\ref{dualN}) is to introduce a twisted boundary condition for the vortex fields,
\beq \label{bcN} V_{\alpha}(x^+) = e^{2 \pi i Q/N} V_{\alpha}(x^-), \quad \mathrm{for}\,\, x \in {\cal S}\eeq
where $Q$ is the charge of our Wilson line. The physical origin of the factor $1/N$ is the fractional charge $2 \pi/N$ of the vortex fields $V_{\alpha}$ under the flux symmetry. Thus, we come to the amazing conclusion that the universal physics in the planar model is periodic in the charge $Q$ of the Wilson line, with period $Q = N$. This is a generalization of the $Q = 1$ periodicity of single flavour QED discussed before. 

Similarly to the $N = 1$ case, a consequence of the $Q \mod N$ periodicity is the emergence of the $C$ and $CP$ symmetries for half-integer values of $Q/N$. As before, this periodicity is due to the identification of $C$, $CP$ conjugate points $Q/N = \pm 1/2$. 

\section{Discussion}\label{sec:Discussion}
We have argued above that non-compact QED with $N$ identical flavours and a $U(1)^{N-1}$ global symmetry acquires a $Q \mod N$ periodicity in the charge $Q$ of the Wilson loop near its
critical point. In the particular case of $N = 1$, Abelian Higgs model, the period is $Q = 1$. It is important to note that this periodicity is emergent, rather than fundamental. On shortest distance scales, QED in three dimensions is perturbative; the electric field produced by an external charge $Q$ is simply Coulombic, $E = \frac{Q\, e^2}{2 \pi r}$, which is obviously not periodic in $Q$. 

One also should not confuse the emergent periodicity with trivial screening of external charges by dynamical fields. Such screening generically takes place in low-dimensional abelian gauge theories and is due to the confining nature of the Coulomb potential (linear in one spatial dimension and logarithmic in two), which leads to binding of a dynamical particle by the external charge\footnote{In two dimensional theories, an infinitely large Wilson loop of charge $Q$ is equivalent to a non-zero value of the topological $\theta$ angle, $\theta = 2 \pi Q$. So the large-distance periodicity in $Q$ is synonymous to periodicity in $\theta$.}. However, this trivial screening typically i) occurs on distance scales $r \gg \xi$, where $\xi = m^{-1}$ is the correlation length of the theory, ii) leads to a resulting period of $Q = 1$. One classic example of this phenomenon is the $Q \mod 1$ periodicity of string tension in massive Schwinger model\cite{Coleman}. The phenomenon considered in the present paper is clearly different since i) the screening occurs on distance scale $r \ll \xi$ (for strong coupling, $e^2 \sim \Lambda$, the screening is actually expected to take place on short-distance cutoff scale), ii) the resulting period is $Q = N$ rather than $Q = 1$. For $N > 1$, in the Coulomb phase of the $U(1)^{N-1}$ symmetric theory, we also expect the usual screening on distances $r \gg \xi$ with period $Q = 1$. 

Physically, the periodicity discussed in this paper is due to the fact that the degrees of freedom responsible for the phase-transition are local fields carrying a fixed magnetic flux, $\Phi_0 = 2 \pi/N$. The Wilson loop of charge $Q$ is then expressed through the linking number $n$ of the worldlines of dual particles with the contour of the loop,
\beq \label{linking} W({\cal C}) = e^{i Q \Phi_0 n} \eeq
and is trivial for $Q = 0 \mod N$. For the special case of $N = 1$ this phenomenon is a manifestation of the fact that the phase transition is driven by magnetic rather than electric degrees of freedom.

The results presented in this paper can be explicitly checked by lattice simulations. The simplest lattice counterparts of the continuum theories under consideration were actually discussed in this paper in sections \ref{sec:LatticeDual} and in the appendix. These lattice theories were previously simulated in a number of studies\cite{Halpern,Laine,mv,SudboBabaev}. We have predicted that near the phase transition all physical observables become periodic in $Q$. However, we have not discussed specific observables. One observable that we suggest for lattice simulations is the electric field produced by a straight temporal Wilson line (we define the electric field $E_i = F_{i3} = - \epsilon_{ij} F_j$). At the critical point, the electric field must have the form,
\begin{figure}[t]
\begin{center}
\includegraphics[angle=-90,width = 0.8\textwidth]{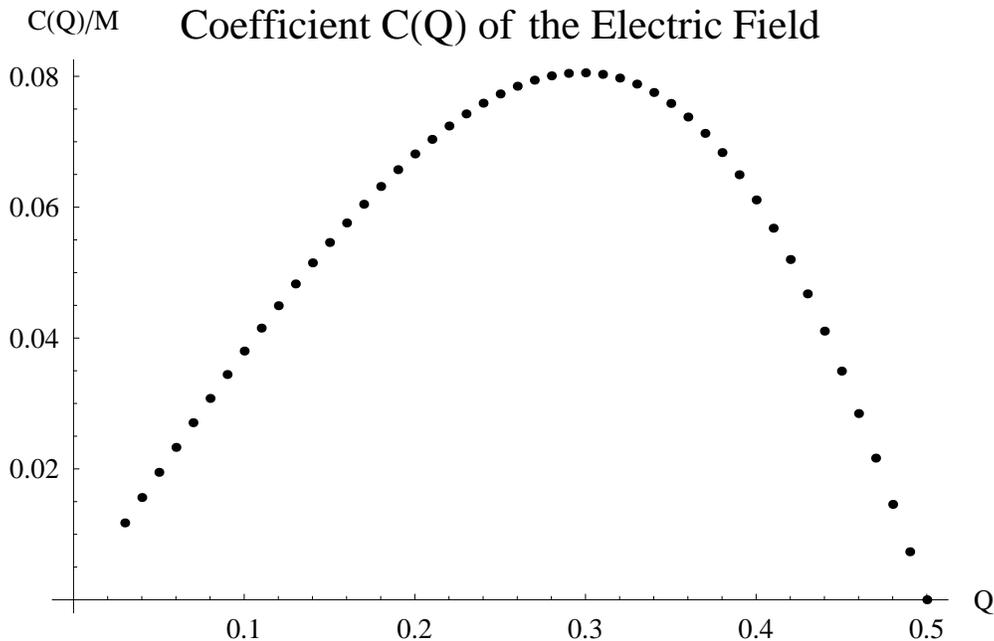}
\caption{Coefficient $C(Q)$ of the electric field at the critical point of the Abelian Higgs model, see eq. (\ref{FieldCrit}), computed from a $1/M$ expansion of the dual theory.} \label{CQ}
\end{center}
\end{figure}
\beq \label{FieldCrit} \langle - i E_r \rangle = C(Q) \frac{1}{r^2}\eeq
This form is dictated by the fact that the magnetic field $-i F_{\mu}$ is a conserved current, which receives no renormalization and hence has conformal dimension $D -1 = 2$. The coefficient $C(Q)$ is a universal function depending only on the charge $Q$ of the Wilson loop\footnote{$C(Q)$ is a real valued function: the electric field is imaginary as we are working in Euclidean space.}. We predict that $C(Q)$ must be periodic with period $Q = N$. In particular, $C(Q)$ vanishes for $Q = 0 \mod N$. Moreover, due to the emergence of $C$ and $CP$ symmetries at half-integer $Q/N$, $C(Q)$ actually vanishes for both integer and half-integer $Q/N$. This does not mean that the electric field vanishes at these special points.  Rather it will be controlled by irrelevant couplings, e.g. one in eq. (\ref{vortexdens}), and will fall off as some higher power of $1/r$ with a non-universal coefficient.

The present paper has concentrated on incorporating Wilson loops into the dual Lagrangian and discussing their general properties. Nevertheless, the dual Lagrangian can also be used for explicit calculations of properties of Wilson loops. Of course, the dual theory is still strongly coupled in the infra-red: we have mapped one difficult problem onto another. However, in the case of the Abelian Higgs model ($N$ = 1), the dual theory is just the three dimensional $U(1)$ symmetric scalar field theory rather than a gauge theory. A wealth of numerical and analytical information is known about the phase transition in this theory (XY universality class). It is known that $\epsilon$-expansion and large-$M$ expansion (whereby the dual field $V$ is promoted to have $M$ components), produce accurate results in this theory. This is in contrast to the $\epsilon$ expansion in the direct theory, which predicts the existence of a critical point only for $N > 182$ (see Ref. \cite{HLM}) (here the field $z$ is promoted to an $N$ component $SU(N)$ multiplet). Likewise the large-$N$ expansion of the direct theory when extrapolated to $N = 1$ is known to produce results for anomalous dimensions, which are numerically notoriously inaccurate. Moreover, we shall argue below that the $SU(N)$ symmetric theory actually does not possess any periodicity in the charge $Q$ of the Wilson line and hence does not capture the qualitative features of the $N = 1$ theory. Thus, at least for the $N = 1$ case, there are clear advantages of performing calculations in the dual, rather than direct theory. In Ref. \onlinecite{MSVBS}, we have used the large-$M$ expansion of the dual theory to explicitly compute the universal function $C(Q)$ of eq. (\ref{FieldCrit}) at $M = \infty$, see Fig. \ref{CQ}. The coefficient $A$ of the linear term of the expansion of $C(Q) \approx A\, Q$ for $Q \to 0$ is actually related to the conductivity in the $XY$ model and is known numerically from Monte-Carlo simulations $A \approx 0.29$, as well as to $O(\epsilon^2)$ in $\epsilon$ expansion $A \approx 0.32$, and to next to leading order in  $1/M$ expansion $A \approx 0.25$ (see \cite{MSVBS} and references therein).

For $U(1)^{N-1}$ symmetric theories with $N > 1$ the dual theory is a gauge theory with $N-1$ gauge fields (the $N = 2$ theory is actually self-dual). Thus, explicit calculations in the dual theory are unlikely to be numerically accurate. However, they may illuminate general features, which are not immediately obvious in the direct theory.

\section{Wilson Loops in the $SU(N)$ symmetric theory}\label{sec:SUN}

It is interesting to ask whether the periodicity in the charge $Q$ of the Wilson loop generalizes to theories with $N$ flavours and a full $SU(N)$ global symmetry. We shall argue below that the answer to this question is no. 

A powerful tool in the analysis of $SU(N)$ symmetric model is the $1/N$ expansion, which allows one to study the physics directly without performing any duality transformations. The $1/N$ expansion is typically performed in the limit $e^2 \to \infty$, so that the bare kinetic term for the gauge field is absent. One also usually replaces the short range repulsive interaction between scalar fields by a hard-constraint, $\sum_{\alpha} z^{\dagger}_{\alpha} z_{\alpha} = \frac{1}{g}$, which can be enforced by a local Lagrange multiplier $\lambda$. One then obtains the Lagrangian of the $CP^{N-1}$ model,
\beq L = |(\d_{\mu} - i A_{\mu}) z|^2 + i \lambda (|z|^2 - \frac{1}{g})\eeq
The behaviour of Wilson loops with $Q \sim O(1)$ in $N$ is easily captured by the $1/N$ expansion. For instance, at the critical point of the theory one recovers the form of the electric field (\ref{FieldCrit}), where to leading order in $1/N$, $C(Q) = \frac{8 Q}{\pi N}$. The $SU(N)$ symmetric theory, thus, clearly does not display a $Q \mod 1$ periodicity of the single flavour theory. 
One might not be too surprised by this fact, since we already saw that the $N$-flavour theory with a $U(1)^{N-1}$ symmetry has only a $Q \mod N$ periodicity. Does the $SU(N)$ symmetric theory share this periodicity of its planar counterpart?

To answer this question we must take $Q \sim O(N)$ in the large $N$ expansion. In this limit, the Wilson loop will modify the saddle-point of the expansion. For $N = \infty$, at the critical point, we should no longer expand around $A_{\mu} = 0, \lambda = 0$, but rather around space-time dependent $A_{\mu}$ and $\lambda$.  By dimensional analysis, a straight temporal Wilson line placed at the origin will generate saddle point fields,
\beq \label{saddlefields} i A_\mu(r)  = \delta_{\mu 0} \frac{a(Q)}{r}, \quad i \lambda(r) = \frac{b(Q)}{r^2}\eeq
The coefficients $a(Q)$ and $b(Q)$ should be chosen in such a way that saddle-point equations are satisfied,
\bea \label{saddleA} \langle J_{\mu}(x)\rangle  &=& \langle z^{\dagger}\overleftrightarrow{D}_{\mu}z(x)\rangle = - Q \delta_{\mu 0} \delta^2(\vec{x})  \\
\label{saddlelambda}\langle z^{\dagger} z(x)\rangle &=& \frac{1}{g}\eea
The expectation values in (\ref{saddleA}), (\ref{saddlelambda}) are to be computed by finding the propagator of the $z$ fields in the background of the saddle point fields (\ref{saddlefields}). We have not been able to find this propagator explicitly. Nevertheless, the saddle-point equation (\ref{saddleA}) is not periodic in $Q/N$, thus, we conclude that the $SU(N)$ symmetric theory is not periodic in $Q$.

One may ask, what makes the $SU(N)$ symmetric theory so different from its deformation with a $U(1)^{N-1}$ symmetry. Our results regarding the planar theory relied on rewriting the problem in terms of dual degrees of freedom. Such a duality transformation, either in the continuum, or on the lattice is, so far, not known in the $SU(N)$ symmetric theory. Nevertheless, we would like to provide some speculations on the possible form of the dual theory and implications for the properties of Wilson loops. 

To construct the dual theory, we first need to identify the dual degrees of freedom. For the planar theories considered above these are vortices: pointlike topological defects of the two dimensional reduction of the theory. Then the three dimensional theory is formulated in terms of vortex loops, i.e. worldlines of the two-dimensional defects. We would like to follow the same procedure for the $SU(N)$ symmetric theory: as a first step we need to find two dimensional topological defects. Fortunately, classical defects of the two dimensional $CP^{N-1}$ model are the very-well known instantons\cite{AVL}. These instantons carry magnetic flux $\Phi = 2 \pi q$, where $q$ is the integer topological charge of the instanton. To make further connection with the planar $N$-flavour model we recall that all the solutions with topological charge $q$ are known exactly, and can be parameterized in terms of $N q$ complex coordinates, $a_{\alpha i}$, with $i = 1 .. q$. For $q > 0$,
\bea \label{instquark} w_{\alpha}(s) &=& c_{\alpha} \prod_{i=1}^q (s-a_{\alpha \,i})\\
z_{\alpha}(s) &=& \frac{w_{\alpha}(s)}{(w^{\dagger} w(s))^{\frac{1}{2}}}\eea
where $s = x_1 + i x_2$. The coefficients $c_{\alpha}$ specify the overall orientation of the instanton in flavour space at infinity and are not very important. The corresponding expressions for $q < 0$ can be obtained by taking $s \to \bar{s}$. We see that the variables $a_{\alpha i}$ are locations of vortices of fields $z_{\alpha}$. Thus, just as in the case of the planar $N$-flavour model, the instanton (vortex) with flux $2 \pi q$ can be decomposed into $q N$ fractional instantons (vortices) to which we can assign flux $2 \pi/N$.

However, there is one major distinction between the planar and $SU(N)$ symmetric models. For the planar model the magnetic flux is concentrated in the core of the fractional vortices, which are assumed to have some microscopic size. On the other hand, for the $SU(N)$ symmetric model, there is no notion of the core-size since the theory is (classically) conformally invariant. The flux density of a configuration of fractional instantons is not localized near their positions $a_{i \alpha}$, but rather is smeared out in a distribution that depends in some highly complicated manner on the ratios of distances between these constituents. An instructive example is the instanton with charge $q = 1$, where the flux produced by the fractional instantons always clumps together into a rotationally invariant distribution.

Thus, despite their similarity to fractional vortices of the planar model, fractional instantons do not carry a local flux and upon transition to three dimensions cannot be promoted to local fields charged under the $U(1)_\Phi$ symmetry. The expression for the Wilson loop (\ref{linking}) in terms of the linking number of vortices with the loop contour is, therefore, inapplicable for the $SU(N)$ symmetric case, and no periodicity in the charge $Q$ of the Wilson loop appears.

\section{Conclusion}\label{sec:Conclusion}
The purpose of this paper was to incorporate Wilson loops into the dual description of critical non-compact abelian gauge theories in three dimensions. This goal has been achieved for non-compact QED with $N$ flavours of identical scalar fields and a $U(1)^{N-1}$ global symmetry. The Abelian Higgs model corresponds to the $N = 1$ case of our general construction. A remarkable property, which follows from the dual description, is that the universal physics close to the phase transition is periodic in charge $Q$ of the Wilson loop with period $Q = N$. In section \ref{sec:Discussion} we have provided a detailed discussion of this unexpected result; here we repeat a few conclusions. We have argued that the periodicity is emergent at the phase transition rather than fundamental to the theory. We also claim that this behaviour is distinct from trivial screening of electric charge in the Coulomb phase of low-dimensional abelian gauge theories. Moreover, we've argued that this periodicity does not generalize to the theory with a full $SU(N)$ invariance. Thus, any attempt to understand the behaviour of Wilson loops at the phase transition of the Abelian Higgs model through the $1/N$ expansion of its $N$-flavour, $SU(N)$ symmetric counterpart will fail to reproduce this qualitative feature. 

The predictions of the present paper can be explicitly tested by lattice simulations. In section \ref{sec:Discussion}, we have suggested one observable: the electric field produced by a straight temporal Wilson line at the critical point, to test the periodicity discussed above on the lattice.

Finally, the application of results of the present paper to the description of an impurity in a two-dimensional antiferromagnet in the neighbourhood of the quantum phase transition to a valence bond solid (VBS) state is a subject of a separate work \cite{MSVBS}. It has been argued in \cite{MSVBS} that a missing spin impurity induces a vortex in the VBS order parameter. We hope that such vortices will be observed by future STM experiments on cuprate compounds.

\acknowledgments
The present work evolved out of collaboration with Subir Sachdev aimed at understanding impurities in quantum antiferromagnets \cite{MSVBS}. I am grateful to Subir for many useful discussions and comments. I would also like to thank Ariel Zhitnitsky for sharing his expertize on fractional instantons. This work was supported by NSF Grant No.\ DMR-0537077.  

\appendix
\section{Duality in the Planar Theory on the Lattice}

In this appendix, we discuss a lattice counterpart of the conjectured continuum duality discussed in section \ref{sec:EasyPlane}.
We start from the action,
\beq \label{SQEDSourceN} S_{QED} = \frac{1}{2 e^2} \sum_{\bar{j} \mu} (\Box A - 2 \pi r)_{\bar{j} \mu}^2 + \frac{1}{2 g} \sum_{\alpha} \sum_{j \mu} (d \theta^{\alpha} - A - 2 \pi n^{\alpha})^2_{j \mu} + i \sum_{\bar{j}\mu} H_{\bar{j} \mu}(\Box A - 2 \pi r)_{\bar{j}\mu}\eeq
This is a generalization of $U(1)$ lattice theory (\ref{SQEDSource}), in which we have introduced $N$ flavours of dynamical matter fields $e^{i \theta^{\alpha}}$, with corresponding integer valued variables $n^{\alpha}$ that ensure the periodicity of $\theta^{\alpha}$ variables. 
As before, we decouple the kinetic term for the gauge field by introducing an auxiliary field $P_{\bar{j} \mu}$ and we Poisson resum the fields $n^{\alpha}_{j \mu}$ by introducing integer valued variables $J^{\alpha}_{j \mu}$,
\beq S \to \frac{e^2}{2} \sum_{\bar{j} \mu} P^2_{\bar{j} \mu} + \frac{1}{2 g} \sum_{\alpha}\sum_{j \mu} (d \theta^{\alpha} - A - 2 \pi n^{\alpha})^2_{j \mu} + 2 \pi i \sum_{\alpha} \sum_{j \mu} n^{\alpha}_{j \mu} J^{\alpha}_{j \mu} + i \sum_{\bar{j} \mu} (P+H)_{\bar{j} \mu} (\Box A - 2 \pi r)_{\bar{j} \mu}\eeq
Shifting, $2 \pi {n'}^{\alpha} =  2\pi n^{\alpha} + A -d \theta^{\alpha}$,
\beq S \to \frac{e^2}{2} \sum_{\bar{j} \mu} P^2_{\bar{j} \mu} + \frac{1}{2 g}\sum_{\alpha}\sum_{j \mu} (2 \pi {n'}^{\alpha})^2_{j \mu} + i \sum_{\alpha}\sum_{j \mu} (2 \pi {n'}^{\alpha}-A+d \theta^{\alpha})_{j \mu} J^{\alpha}_{j \mu} + i \sum_{\bar{j} \mu} (P + H)_{\bar{j} \mu}(\Box A - 2 \pi r)_{\bar{j} \mu}\eeq
Performing the integral over $\theta^{\alpha}_{j \mu}$ we obtain a set of constraints
\beq \label{mconstrN} \nabla \cdot J^{\alpha} = 0\eeq Physically, $J^{\alpha}_{j \mu}$ represent the world-lines of $e^{i \theta^{\alpha}}$ particles. The singlet combination $\sum_{\alpha} J^{\alpha}$ has long-range $1/r$ Coulomb interaction, while the non-singlet combinations have short range interactions. We solve the constraints (\ref{mconstrN}) in terms of integer valued fields $b^{\alpha}_{\bar j \mu}$,
\beq \label{mconstsolN} J^{\alpha}_{j \mu} = (\Box b^{\alpha})_{j \mu}\eeq Integrating over the ${n'}^{\alpha}$ fields,
\beq S_{QED} \to \frac{e^2}{2} \sum_{\bar{j} \mu} P^2_{\bar{j} \mu} + i \sum_{\bar{j} \mu} \big(\Box(P+H) - \sum_{\alpha}J^{\alpha}\big)_{j \mu} A_{j \mu} - 2 \pi i \sum_{\bar{j} \mu} (P+H)_{\bar{j} \mu} r_{\bar{j} \mu} + \frac{g}{2} \sum_{\alpha}\sum_{j \mu} {J^{\alpha}_{j \mu}}^2\eeq
Performing the integral over the $A$ field, we obtain a constraint,
\beq \label{PconstrN}\Box(P+H)_{j \mu} - \sum_{\alpha} J^{\alpha}_{j \mu} = 0\eeq
Recalling (\ref{mconstsolN}) we can solve (\ref{PconstrN}) by introducing a real field $\varphi$,
\beq P + H - \sum_{\alpha} b^{\alpha} = \frac{d \varphi}{2 \pi}\eeq
arriving at the action,
\beq S \to \frac{e^2}{8 \pi^2} \sum_{{\bar j} \mu}(d \varphi - 2 \pi H + 2 \pi \sum_{\alpha} b^{\alpha})^2_{\bar{j} \mu} + \frac{g}{2} \sum_{\alpha}\sum_{j \mu} (\Box b^{\alpha})^2_{j \mu}-i \sum_{\bar{j}} \varphi_{\bar{j}} s_{\bar{j}}\eeq
We now Poisson resum the field $b^{\alpha}$ by introducing integer valued variables $l^{\alpha}$,
\beq S \to \frac{e^2}{8 \pi^2} \sum_{{\bar j} \mu}(d \varphi - 2 \pi H + 2 \pi \sum_{\alpha} b^{\alpha})^2_{\bar{j} \mu} + \frac{g}{2} \sum_{\alpha}\sum_{j \mu} (\Box b^{\alpha})^2_{j \mu}-i \sum_{\bar{j}} \varphi_{\bar{j}} s_{\bar{j}} + 2 \pi i \sum_{\alpha} \sum_{\bar{j} \mu} l^{\alpha}_{\bar{j} \mu} b^{\alpha}_{\bar{j} \mu}\eeq
Now $b^{\alpha}$ are not constrained and it is convenient to change variables to singlet and non-singlet components $b^0$ and $b^{a}$,
\beq b^{\alpha} = b^0 t^0_{\alpha} + b^a t^a_{\alpha}\eeq
where $t^0$ and $t^a$, $a = 1..N-1$, are normalized generators of singlet and non-singlet symmetry groups respectively, i.e. $t^0_{\alpha} = \frac{1}{\sqrt{N}}$, $\sum_{\alpha} t^a_{\alpha} = 0$ and $\sum_{\alpha} t^{a}_{\alpha} t^{b}_{\alpha} = \delta^{a b}$. We also introduce singlet and  non-singlet combinations,
\beq \label{lrot} l^0 = \frac{1}{\sqrt{N}} \sum_{\alpha} l^{\alpha},\quad l^a = \sum_{\alpha} l^{\alpha} t^a_{\alpha}\eeq
Note that $l^0$ and $l^{a}$ are generally not integer valued. Hence, our action becomes,
\bea S &\to& \frac{e^2}{8 \pi^2} \sum_{\bar{j} \mu} (d \varphi - 2 \pi H + 2 \pi \sqrt{N} b^0)^2_{\bar{j} \mu} + \frac{g}{2} \sum_{j \mu} (\Box b^0)^2_{j \mu} + 2\pi i \sum_{\bar{j} \mu} l^0_{\bar{j} \mu} b^0_{\bar{j} \mu} - i \sum_{\bar{j}} \varphi_{\bar{j}} s_{\bar{j}}\nn \\&+&\frac{g}{2}\sum_a \sum_{j \mu} (\Box b^a)^2_{j \mu} + 2 \pi i \sum_a \sum_{\bar{j} \mu} l^a_{\bar{j} \mu} b^a_{\bar{j} \mu}
\eea
Shifting $2 \pi \sqrt{N} {b'}^0 = d \varphi + 2 \pi \sqrt{N} b^0$, and integrating over $\varphi$, we obtain,
\bea \label{SdualNsingfl}S &\to& \frac{e^2 N}{2} \sum_{\bar{j} \mu} ({b'}^0 - \frac{1}{\sqrt{N}} H)^2_{\bar{j} \mu} + \frac{g}{2} \sum_{j \mu} (\Box {b'}^0)^2_{j \mu} + 2\pi i \sum_{\bar{j} \mu} l^0_{\bar{j} \mu} {b'}^0_{\bar{j} \mu} \\ &+& \frac{g}{2}\sum_a \sum_{j \mu} (\Box b^a)^2_{j \mu} + 2 \pi i \sum_a \sum_{\bar{j} \mu} l^a_{\bar{j} \mu} b^a_{\bar{j} \mu}
\eea
with the constraint,
\beq \label{constrsing}(\nabla \cdot l^0)_{\bar{j}} = \sqrt{N} s_{\bar{j}}\eeq
Moreover, performing a shift $b^{a} \to b^{a} + d \phi$ in (\ref{SdualNsingfl}) we learn that the flavoured components must satisfy,
\beq \label{constrflav}\nabla \cdot l^a = 0\eeq
The constraints (\ref{constrsing}) and (\ref{constrflav}) can be combined to give,
\beq \label{constralpha}(\nabla \cdot l^{\alpha})_{\bar{j}} =  s_{\bar{j}}\quad \mathrm{for}\, \mathrm{all}\, \alpha\eeq
Physically, $l^{\alpha}$ corresponds to worldlines of vortices of $e^{i \theta_{\alpha}}$ field. The singlet component of this field $l^0$ corresponds to local vortices, when $e^{i \theta_{\alpha}}$ have the same winding for all $\alpha$.
Now, we can integrate the fields ${b^0}'$ and $b^a$ out in eq. (\ref{SdualNsingfl}),
\bea \label{Nlatl1}S &=& \frac{1}{2} \sum_{j j' \mu} (2 \pi l^0)_{\bar{j} \mu} D^S_{jj'}(2 \pi l^0)_{\bar{j'} \mu} + 2 \pi i \sum_{j j' \mu} H_{\bar{j} \mu} (e^2 D^S_{j j'}) (\sqrt{N} l^0)_{\bar{j'} \mu} \\\label{Nlatl2}&+&  \frac{g}{2 e^2} \sum_{j j'} s_j D^S_{j j'} s_{j'}+ \frac{e^2 g}{2} \sum_{j j'\mu} (\Box H)_{j \mu} D^S_{j j'} (\Box H)_{j' \mu} - 2 \pi i g \sum_{\bar{j} \mu} H_{\bar{j} \mu} (d D^S s)_{\bar{j} \mu}\\&+&
\label{Nlatl3}\frac{1}{2}\sum_a  \sum_{j j' \mu} (2 \pi l^a)_{\bar{j} \mu} D^f_{jj'}(2 \pi l^a)_{\bar{j'} \mu}\eea
with the kernels,
\bea \label{Ds} D^S_{j j'} &=& \frac{1}{V} \sum_k \frac{1}{e^2 N +g\sum_{\nu} 4 \sin^2{\frac{k_{\nu} a}{2}}}e^{i k (j-j')}\\
\label{Df}D^f_{j j'} &=& \frac{1}{V} \sum_k \frac{1}{g\sum_{\nu} 4 \sin^2{\frac{k_{\nu} a}{2}}}e^{i k (j-j')}\eea
The first term in (\ref{Nlatl1}) corresponds to a short range interaction between singlet components of vortices. This is in agreement with the expectation that local vortices should have local interactions. On the other hand, as we see from (\ref{Nlatl3}), non-singlet components of $l$, i.e. global vortices, have long-range interactions. 

Now let's see how our vortices couple to the magnetic source field $H_{j \mu}$. This coupling is given by the second term in (\ref{Nlatl1}).  Rewriting,
\beq i \Phi = 2 \pi i \sum_{j j' \mu} H_{\bar{j} \mu} (e^2 D^S_{j j'}) (\sqrt{N} l^0)_{\bar{j'} \mu} = 2 \pi i \sum_{j j' \mu} H_{\bar{j} \mu} (e^2 D^S_{j j'}) (\sum_{\alpha} l^{\alpha})_{\bar{j'} \mu}\eeq we observe that each vortex $l^{\alpha}$ carries a flux equal to $2 \pi e^2 D^S(k=0) = 2 \pi/N$. This is again in accordance with our arguments in section \ref{sec:EasyPlane}.

Finally, for completeness we also discuss the terms in eq. (\ref{Nlatl2}). These terms are not very important as they don't couple to the dynamical variables $l^{\alpha}$. The first term in (\ref{Nlatl2}) corresponds to bare, short range, interaction between monopoles, while the third term corresponds to the monopole's contribution to the magnetic flux.

Now, we would like to rewrite the action (\ref{Nlatl1}) in a local form. We consider the action,
\bea \label{U1locN} S &=& \frac{1}{2 \tilde{e}^2} \sum_a \sum_{j \mu} (\Box B^a)^2_{j \mu} + \frac{1}{2 \tilde{g}} \sum_{\alpha} \sum_{\bar{j} \mu} (d \varphi^{\alpha}-B^a t^a_{\alpha} - \frac{2 \pi}{N} H+ 2 \pi m^{\alpha})^2_{{\bar{j} \mu}} \\ \label{U1locN2}&+&
\frac{t}{2} \sum_{j {\mu}} \left(\sum_{\alpha} \Box m^{\alpha}\right)^2_{j {\mu}} - i \sum_{\bar{j}} \left(\sum_{\alpha} \varphi^{\alpha}_{\bar{j}}\right) s_{\bar{j}}\eea
The first line (\ref{U1locN}) is the simplest lattice generalization of the continuum action (\ref{dualN}). We have a set of $N-1$ dynamical gauge field $B^a$ which couple to non-singlet combinations of currents associated with a set of $N$ matter fields $V_{\alpha} \sim e^{i \varphi^{\alpha}}$. The integer variables $m^{\alpha}_{j \mu}$ ensure the periodicity of $\varphi^{\alpha}$. In the second line (\ref{U1locN2}) we introduce the monopole source field $s$ which couples to the combination $\sum_{\alpha} \varphi^{\alpha}$, in accordance with our continuum guess (\ref{VN}). Also, observe the coefficient $2 \pi/N$, corresponding to the flux of a single vortex, in the coupling to the source field $H$.

We have also introduced an additional term with coupling constant $t$ in eq. (\ref{U1locN2}). This term gives local interactions to the vortices of $e^{i\varphi^{\alpha}}$ fields. We don't actually expect this term to drastically modify the critical properties of our theory (as we shall see, it will change some ultra-local interactions into just local ones), nevertheless, we have included it to make the similarity with the direct theory more pronounced. 

We would like to make the connection between actions (\ref{U1locN}) and (\ref{SQEDSourceN}). As usual, we Poisson resum $m^{\alpha}$ in (\ref{U1locN}) with the help of integer valued variables $l^{\alpha}_{\bar{j} \mu}$ and perform a shift, $2 \pi {m'}^{\alpha} = 2 \pi m^{\alpha} - B^a t^a_{\alpha} + d\varphi^{\alpha}$,
\bea  S &=& \frac{1}{2 \tilde{e}^2} \sum_a \sum_{j \mu} (\Box B^a)^2_{j \mu} + \frac{(2 \pi)^2}{2 \tilde{g}} \sum_{\alpha} \sum_{\bar{j} \mu} (m'^{\alpha}- \frac{1}{N} H)^2_{{\bar{j} \mu}} +
\frac{t}{2} \sum_{j {\mu}} \left(\sum_{\alpha} \Box m'^{\alpha}\right)^2_{j {\mu}} \\&-& i \sum_{\bar{j}} \left(\sum_{\alpha} \varphi^{\alpha}_{\bar{j}}\right) s_{\bar{j}} + i \sum_{\alpha} \sum_{{\bar j}\mu} l^{\alpha}_{\bar{j} \mu} (2 \pi {m'}^{\alpha} + B^a t^a_{\alpha} -d \varphi^{\alpha})_{\bar{j} \mu}\eea
By integrating over $\varphi^{\alpha}$ we recover the constraint (\ref{constralpha}). Next, we go to the rotated variables (\ref{lrot}) and perform analogous rotation on on ${m'}^\alpha$. Then,
\bea  S &=&  \frac{(2 \pi)^2}{2 \tilde{g}} \sum_{\bar{j} \mu} ({m'}^{0}- \frac{1}{\sqrt{N}} H)^2_{{\bar{j} \mu}} +
\frac{t N}{2} \sum_{j {\mu}}  (\Box {m'}^{0})^2_{j {\mu}} + 2 \pi i \sum_{\bar{j} \mu} l^0_{\bar{j} \mu} {m'}^0_{\bar{j}{\mu}}\\&+&\frac{1}{2 \tilde{e}^2} \sum_a \sum_{j \mu} (\Box B^a)^2_{j \mu}  +  \frac{(2 \pi)^2}{2 \tilde{g}} \sum_a \sum_{\bar{j} \mu} ({m'}^{a})^2_{{\bar{j} \mu}}+i \sum_{a} \sum_{{\bar j}\mu} l^{a}_{\bar{j} \mu} (2 \pi {m'}^{a} + B^a)_{\bar{j} \mu}\eea
Integrating over ${m'}^a$ we obtain,
\bea  S &=&  \frac{(2 \pi)^2}{2 \tilde{g}} \sum_{\bar{j} \mu} ({m'}^{0}- \frac{1}{\sqrt{N}} H)^2_{{\bar{j} \mu}} +
\frac{t N}{2} \sum_{j {\mu}}  (\Box {m'}^{0})^2_{j {\mu}} + 2 \pi i \sum_{\bar{j} \mu} l^0_{\bar{j} \mu} {m'}^0_{\bar{j}{\mu}}\\&+&\frac{1}{2 \tilde{e}^2} \sum_a \sum_{j \mu} (\Box B^a)^2_{j \mu}  +  +i \sum_{a} \sum_{{\bar j}\mu} l^{a}_{\bar{j} \mu} B^a_{\bar{j} \mu} + \frac{\tilde{g}}{2 }\sum_{a} \sum_{{\bar j}\mu} (l^a)^2_{\bar{j} \mu}\label{U1locNfin}\eea
We note that the above action, except for the last term in eq. (\ref{U1locNfin}), is exactly the same as that in (\ref{SdualNsingfl}) with the identification, ${m'}^0 = {b'}^0$, $B^a = 2 \pi b^a$, $\tilde{e}^2 = \frac{(2 \pi)^2}{g}$, $\tilde{g} = \frac{(2 \pi)^2}{e^2 N}$, $t = g/N$. As for the last term $\frac{\tilde{g}}{2 }\sum_{a} \sum_{{\bar j}\mu} (l^a)^2_{\bar{j} \mu}$, it gives rise to an ultra-local interaction between global vortices. Since these vortices already interact with a long-range potential (\ref{Df}), we don't expect this term to alter the critical properties of the theory. We also note that the somewhat unusual term with the coefficient $t$ in (\ref{U1locN2}) is related to the coupling $g$ in the singlet kernel $D^S$ (\ref{Ds}) (but not in the flavoured kernel $D^f$ (\ref{Df})!). Setting $t = 0$, would make interactions between singlet vortices, ultra-local, instead of just short-range. Again, we don't expect such a change to alter the critical properties of the theory.

Thus, we have argued that the lattice actions (\ref{U1locN}) and (\ref{SQEDSourceN}) are equivalent, up to local interactions in the flavoured sector.

\end{document}